\documentclass[letterpaper,twocolumn,prl,aps,superscriptaddress,amsmath,amssymb,floatfix]{revtex4-1}
\usepackage{mathptmx}
\usepackage[latin9]{inputenc}
\usepackage{color}
\usepackage{amsmath}
\usepackage{amssymb}
\usepackage{graphicx}
\usepackage{esint}
\usepackage[unicode=true,
 bookmarks=true,bookmarksnumbered=false,bookmarksopen=false,
 breaklinks=false,pdfborder={0 0 1},backref=false,colorlinks=true]
 {hyperref}
\hypersetup{
 linkcolor=magenta,urlcolor=blue,citecolor=blue,pdfstartview={FitH},hyperfootnotes=false}

\makeatletter

\newcommand{\lyxmathsym}[1]{\ifmmode\begingroup\def\b@ld{bold}
  \text{\ifx\math@version\b@ld\bfseries\fi#1}\endgroup\else#1\fi}

\usepackage{textcomp}
\usepackage{epstopdf}

\usepackage{amsfonts}

\pdfpageheight\paperheight
\pdfpagewidth\paperwidth

\@ifundefined{textcolor}{}{%
 \definecolor{BLACK}{gray}{0}
 \definecolor{WHITE}{gray}{1}
 \definecolor{RED}{rgb}{1,0,0}
 \definecolor{GREEN}{rgb}{0,1,0}
 \definecolor{BLUE}{rgb}{0,0,1}
 \definecolor{CYAN}{cmyk}{1,0,0,0}
 \definecolor{MAGENTA}{cmyk}{0,1,0,0}
 \definecolor{YELLOW}{cmyk}{0,0,1,0}
}

\usepackage{xcolor}\usepackage{soul}
\definecolor{blue}{rgb}{0,0,1}
\definecolor{red}{rgb}{1,0,0}
\definecolor{green}{rgb}{0,1,0}

\makeatother

\begin{document}
\affiliation{Key Laboratory of Quantum Information, Chinese Academy of Sciences,
University of Science and Technology of China, Hefei 230026, P. R. China.}
\affiliation{Hefei National Laboratory for Physical Sciences at the Microscale,
University of Science and Technology of China, Hefei, Anhui, China,
230026}
\affiliation{Department of Electrical Engineering, Yale University, New Haven,
CT 06511, USA}
\affiliation{CAS Center For Excellence in Quantum Information and Quantum Physics,
University of Science and Technology of China, Hefei, Anhui 230026,
P. R. China.}
\title{Planar-integrated magneto-optical trap}
\author{Liang~Chen}
\affiliation{Key Laboratory of Quantum Information, Chinese Academy of Sciences,
University of Science and Technology of China, Hefei 230026, P. R. China.}
\affiliation{CAS Center For Excellence in Quantum Information and Quantum Physics,
University of Science and Technology of China, Hefei, Anhui 230026,
P. R. China.}
\author{Chang-Jiang~Huang}
\affiliation{Key Laboratory of Quantum Information, Chinese Academy of Sciences,
University of Science and Technology of China, Hefei 230026, P. R. China.}
\affiliation{CAS Center For Excellence in Quantum Information and Quantum Physics,
University of Science and Technology of China, Hefei, Anhui 230026,
P. R. China.}
\author{Xin-Biao~Xu}
\affiliation{Key Laboratory of Quantum Information, Chinese Academy of Sciences,
University of Science and Technology of China, Hefei 230026, P. R. China.}
\affiliation{CAS Center For Excellence in Quantum Information and Quantum Physics,
University of Science and Technology of China, Hefei, Anhui 230026,
P. R. China.}
\author{Zheng-Tian~Lu}
\affiliation{Hefei National Laboratory for Physical Sciences at the Microscale,
University of Science and Technology of China, Hefei, Anhui, China,
230026}
\affiliation{CAS Center For Excellence in Quantum Information and Quantum Physics,
University of Science and Technology of China, Hefei, Anhui 230026,
P. R. China.}
\author{Zhu-Bo~Wang}
\affiliation{Key Laboratory of Quantum Information, Chinese Academy of Sciences,
University of Science and Technology of China, Hefei 230026, P. R. China.}
\affiliation{CAS Center For Excellence in Quantum Information and Quantum Physics,
University of Science and Technology of China, Hefei, Anhui 230026,
P. R. China.}
\author{Guang-Jie~Chen}
\affiliation{Key Laboratory of Quantum Information, Chinese Academy of Sciences,
University of Science and Technology of China, Hefei 230026, P. R. China.}
\affiliation{CAS Center For Excellence in Quantum Information and Quantum Physics,
University of Science and Technology of China, Hefei, Anhui 230026,
P. R. China.}
\author{Ji-Zhe~Zhang}
\affiliation{Key Laboratory of Quantum Information, Chinese Academy of Sciences,
University of Science and Technology of China, Hefei 230026, P. R. China.}
\affiliation{CAS Center For Excellence in Quantum Information and Quantum Physics,
University of Science and Technology of China, Hefei, Anhui 230026,
P. R. China.}
\author{Hong X.~Tang}
\affiliation{Department of Electrical Engineering, Yale University, New Haven,
CT 06511, USA}
\author{Chun-Hua~Dong}
\affiliation{Key Laboratory of Quantum Information, Chinese Academy of Sciences,
University of Science and Technology of China, Hefei 230026, P. R. China.}
\affiliation{CAS Center For Excellence in Quantum Information and Quantum Physics,
University of Science and Technology of China, Hefei, Anhui 230026,
P. R. China.}
\author{Wen~Liu}
\affiliation{USTC Center for Micro-and Nanoscale Research and Fabrication, University
of Science and Technology of China, Hefei, Anhui 230026, P. R. China.}
\author{Guo-Yong~Xiang}
\affiliation{Key Laboratory of Quantum Information, Chinese Academy of Sciences,
University of Science and Technology of China, Hefei 230026, P. R. China.}
\affiliation{CAS Center For Excellence in Quantum Information and Quantum Physics,
University of Science and Technology of China, Hefei, Anhui 230026,
P. R. China.}
\author{Guang-Can~Guo}
\affiliation{Key Laboratory of Quantum Information, Chinese Academy of Sciences,
University of Science and Technology of China, Hefei 230026, P. R. China.}
\affiliation{CAS Center For Excellence in Quantum Information and Quantum Physics,
University of Science and Technology of China, Hefei, Anhui 230026,
P. R. China.}
\author{Chang-Ling~Zou}
\email{clzou321@ustc.edu.cn}

\affiliation{Key Laboratory of Quantum Information, Chinese Academy of Sciences,
University of Science and Technology of China, Hefei 230026, P. R. China.}
\affiliation{CAS Center For Excellence in Quantum Information and Quantum Physics,
University of Science and Technology of China, Hefei, Anhui 230026,
P. R. China.}
\date{\today}
\begin{abstract}
The magneto-optical trap (MOT) is an essential tool for collecting
and preparing cold atoms with a wide range of applications. We demonstrate
a planar-integrated MOT by combining an optical grating chip with
a magnetic coil chip. The flat grating chip simplifies the conventional
six-beam configuration down to a single laser beam; the flat coil
chip replaces the conventional anti-Helmholtz coils of a cylindrical
geometry. We trap $10^{4}$ cold $^{87}\text{Rb}$ atoms in the planar-integrated
MOT, at a point 3 -- 9 mm above the chip surface. This novel configuration
effectively reduces the volume, weight, and complexity of the MOT,
bringing benefits to applications including gravimeter, clock and
quantum memory devices.
\end{abstract}
\maketitle

\section{Introduction}

The magneto-optical trap (MOT) is one of the most important experimental
platforms in atomic physics~\citep{Zhai2015,Firstenberg2016,Cooper2019,Tomza2019}.
Cold atoms prepared by a MOT are widely used in quantum measurement
and metrology applications~\citep{Liu2018,Grotti2018,Udem2002}.
For example, the atomic gravimetry with precision achieving $\text{\ensuremath{\mu}Gal/\ensuremath{\sqrt{\text{Hz}}}}$
has been demonstrated~\citep{Hu2013}. The conventional MOT apparatus
consists of three orthogonal pairs of retro-reflected laser beams
and a pair of anti-Helmholtz coils in a cylindrical geometry~\citep{Raab1987}.

Over the past decade, great efforts have been devoted to minimizing
the MOT system. Most attentions are paid to reducing the bulky optical
system. For example, by using a pyramidal retroreflector, only one
incident laser beam is needed~\citep{Lee1996,Vangeleyn2009,Pollock2009,Pollock2011}.
More recently, the idea was further developed by replacing the pyramidal
retroreflector with a completely flat chip consisting of three gratings,
demonstrating a MOT that can capture as many as $10^{7}$ atoms with
a single incident laser beam~\citep{Vangeleyn2010,Nshii2013,Lee2013,McGilligan2016,Cotter2016}.
In contrast, the original bulky anti-Helmholtz coils still remain.
Although U- or Z-shaped wires were employed in an atom chip to assist
the MOT, external Helmholtz coils were still required to provide a
bias field~\citep{Reichel1999,Denschlag1999,Folman2000,Pollock2009,Pollock2011,Rushton2016}.

In this work, we develop a planar-integrated MOT (piMOT) configuration
based on both a grating chip and a planar $3\,\mathrm{cm}\times3\,\mathrm{cm}$
coil chip. The coil chip generates a quadrupole magnetic field several
millimeters above the chip surface, matching the working points of
the optical grating chip. While carrying a current of $0.9\,\mathrm{A}$,
the magnetic field gradient reaches $9.8\,\mathrm{G/cm}$, and the
low working voltage of $2.5\,\mathrm{V}$ and power of $2.5\,\mathrm{W}$
allows it to be powered by batteries. With both chips stacked outside
a glass vacuum cell, we trap $10^{4}$ $^{87}\text{Rb}$ atoms. The
piMOT is simple, portable, and low-cost. It opens the possibility
for further monolithic integration of the cold atom system with photonic
chips~\citep{Xie2019,Hu2020,Wang2021,McGehee2021}, with future applications
including portable gravimeter~\citep{Kasevich1991,Peters1999,Peters2001,Wang2018},
clock~\citep{ESSEN1955,DeBeauvoir1997,Laurent2006,Liu2017,Liu2017-2}
and quantum memory devices~\citep{Duan2001,Ding2015,Wang2019-2,Wen2019}.

\begin{figure*}
\begin{centering}
\includegraphics[width=1\linewidth]{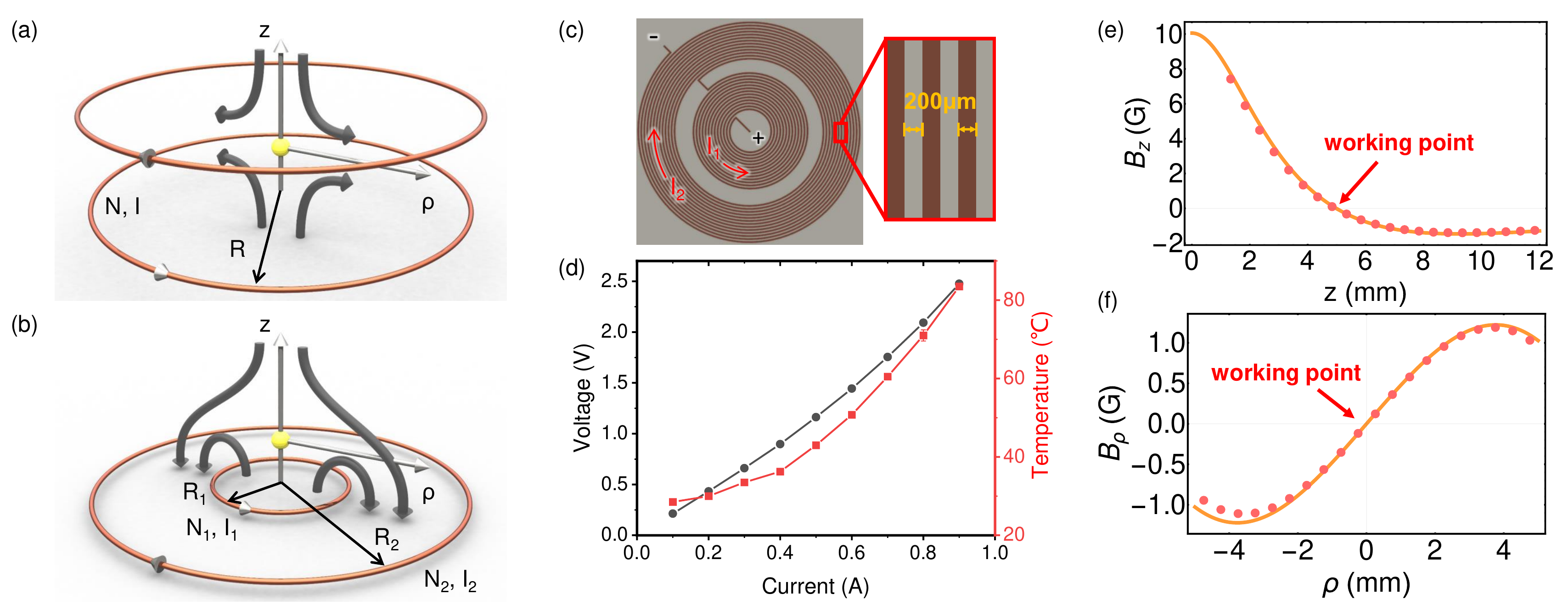}
\par\end{centering}
\caption{The planar quadrupole-field coil chip. (a) The conventional anti-Helmholtz
coils in a three-dimensional cylindrical geometry. (b) The planar
coils. Parameters $R,\,N,\,I$ are the radii, numbers of turns, and
current of each annular coils, respectively. The magnetic field lines
are indicated by the grey curves. (c) A practical coil chip consisting
of embedded spiraling wires. (d) Steady-state voltage and temperature
of the coil chip versus the applied currents. (e) and (f) The magnetic
field components along $z$- and $\rho$-axis.}

\label{Fig1}
\end{figure*}

\section{The coil chip}

The atoms are trapped in the MOT at the working point where the field
equals zero and strong field gradients are present in all directions~\citep{Bergeman1987}.
A conventional MOT generates this magnetic field using a pair of anti-Helmholtz
coils~\citep{Raab1987}, which is composed by a pair of identical
coils (same radius $R$, current $I$, and turns $N$) with opposite
current directions, as shown in Fig.$\,$\ref{Fig1}(a). The working
point of this coil configuration is exactly its geometrical center.
This cylindrical geometry poses limitations on a compact cold-atom
system. For example, the vacuum cell needs to be inserted in between
the pair of coils, thus limiting the minimum size of the coils. In
order to reach the required \textasciitilde{} 10 G/cm field gradient,
the required currents in the coils grow with the cube of the coil
size, which then leads to heat dissipation problems. These limitations
can be overcome with a planar coil design to achieve a compact system.

We propose the coplanar annular coil chip {[}Fig.$\,$\ref{Fig1}(b){]},
consisting of two coaxial coils with different radii $R_{\text{1}}$
and $R_{\text{2}}$ ($R_{\text{1}}<R_{2}$), different numbers of
turns $N_{\text{1}}$ and $N_{\text{2}}$, and opposite currents $I_{1}$
and -$I_{2}$ ($I_{1,2}>0$). Here, we define the center of the coils
as the origin of the circular coordinate system and the $z$ axis
perpendicular to the coil plane. At a point ($\rho=0,\,\varphi,\,z=z_{0}$)
on the $z$ axis, the coils produce a magnetic field intensity $\boldsymbol{B}=\left(B_{\rho},\,B_{\varphi}=0,\,B_{z}\right)$
as~\citep{Bergeman1987}
\begin{align}
B_{z}\left(0,\varphi,z_{0}\right) & =\frac{\mu_{0}}{2}\left(\frac{N_{1}R_{1}^{2}I_{1}}{\left(R_{1}^{2}+z_{0}^{2}\right)^{\frac{3}{2}}}-\frac{N_{2}R_{2}^{2}I_{2}}{\left(R_{2}^{2}+z_{0}^{2}\right)^{\frac{3}{2}}}\right),\label{eq:1}
\end{align}

\begin{align}
B_{\rho}\left(0,\varphi,z_{0}\right) & =0,\label{eq:2}
\end{align}
and the corresponding magnetic field gradients
\begin{align}
\frac{\partial B_{z}}{\partial z}\left(0,\varphi,z_{0}\right) & =-\frac{3\mu_{0}z_{0}}{2}\left(\frac{N_{1}R_{1}^{2}I_{1}}{\left(R_{1}^{2}+z_{0}^{2}\right)^{\frac{5}{2}}}-\frac{N_{2}R_{2}^{2}I_{2}}{\left(R_{2}^{2}+z_{0}^{2}\right)^{\frac{5}{2}}}\right),\label{eq:3}\\
\frac{\partial B_{\rho}}{\partial\rho}\left(0,\varphi,z_{0}\right) & =\frac{3\mu_{0}z_{0}}{4}\left(\frac{N_{1}R_{1}^{2}I_{1}}{\left(R_{1}^{2}+z_{0}^{2}\right)^{\frac{5}{2}}}-\frac{N_{2}R_{2}^{2}I_{2}}{\left(R_{2}^{2}+z_{0}^{2}\right)^{\frac{5}{2}}}\right),\label{eq:4}
\end{align}
Due to the cylindrical symmetry, $\partial B_{\rho}/\partial\rho=-\frac{1}{2}\partial B_{z}/\partial z$
on the $z$-axis. For simplicity, we first consider the case of balanced
currents $I_{1}=I_{2}=I$, which allows the two coils to be connected
in series with one current supply. For appropriate $R_{1}$ and $R_{2}$,
we can adjust $N_{\text{1}}$ and $N_{\text{2}}$ to make the field
intensities zero at a desired height $h$, i.e. at the target MOT
working point $\left(0,0,h\right)$. As a result, this coplanar coil
configuration is able to provide the quadrupole magnetic field for
realizing MOT, with a working point above the surface of the coil
plane.

A coil chip is constructed based on the printed-circuit-board (PCB)
technology, with a chip thickness of $250\,\mathrm{\mu m}$ and an
edge length of $3\,\mathrm{cm}$. The coils are made by copper wire
printed on a square Rogers ceramic substrate {[}Fig.$\,$\ref{Fig1}(c){]}.
The thickness, width, and spacing of the wire are $70\,\mathrm{\mu m}$,
$200\,\mathrm{\mu m}$, and $200\,\mathrm{\mu m}$, respectively.
The inner coil has $N_{1}=13$ turns, with a radius ranging from 2.6~mm
to 7.8~mm, and the outer coil has $N_{2}=13$ turns, with a radius
ranging from 9.6~mm to 14.8~mm. According to Eq.~(\ref{eq:1}-\ref{eq:4}),
this coil chip has a working point at the height $h=5.0\,\mathrm{mm}$,
an axial field gradient of $\partial B_{z}/\partial z=9.8\,\mathrm{G/cm}$
and a radial gradient of $\partial B_{\rho}/\partial\rho=-4.9\,\mathrm{G/cm}$
when carrying a current of $I=0.9\,\mathrm{A}$.

Figures$\,$\ref{Fig1}(e) and (f) show the calculation and measurement
results of the axial and radial magnetic fields ($B_{z}$ and $B_{\rho}$).
As designed, the coil chip provides an axial gradient of $9.8\,\mathrm{G/cm}$
and a radial gradient of $-4.8\,\mathrm{G/cm}$ around the working
point with $h=5.0\,\mathrm{mm}$. From the measured field distribution,
in a range of about $5\,\mathrm{mm}$ along both $\rho$ and $z$
directions around the working point, the coil chip can provide the
desired magnetic field gradients for the MOT. This provides a sufficiently
large volume for confining cold atoms.

Besides the magnetic fields, we also evaluate the performance of the
coil chip for practical applications by measuring the steady-state
voltage and temperature of the chip versus the current $I$ {[}Fig.$\,$\ref{Fig1}(d){]}.
The coil chip is placed in an ambient environment underneath the glass
cell. At the working current I = 0.9 A, the voltage is measured to
be 2.5 V, corresponding to a resistive heating power of 2.2 W. The
chip temperature remains at around $84\,\lyxmathsym{\textcelsius}$.
For application without need for modulation, the coil chip can be
replaced by a ring of permanent magnet to remove the power supply
and avoid heating issue. In conclusion, the coil chip can provide
a stable magnetic field for MOT at a low power setting. This is made
possible by the small distances in the planar integration design.

\section{Experimental performance of the planar-integrated MOT}

\begin{figure}
\begin{centering}
\includegraphics[width=1\linewidth]{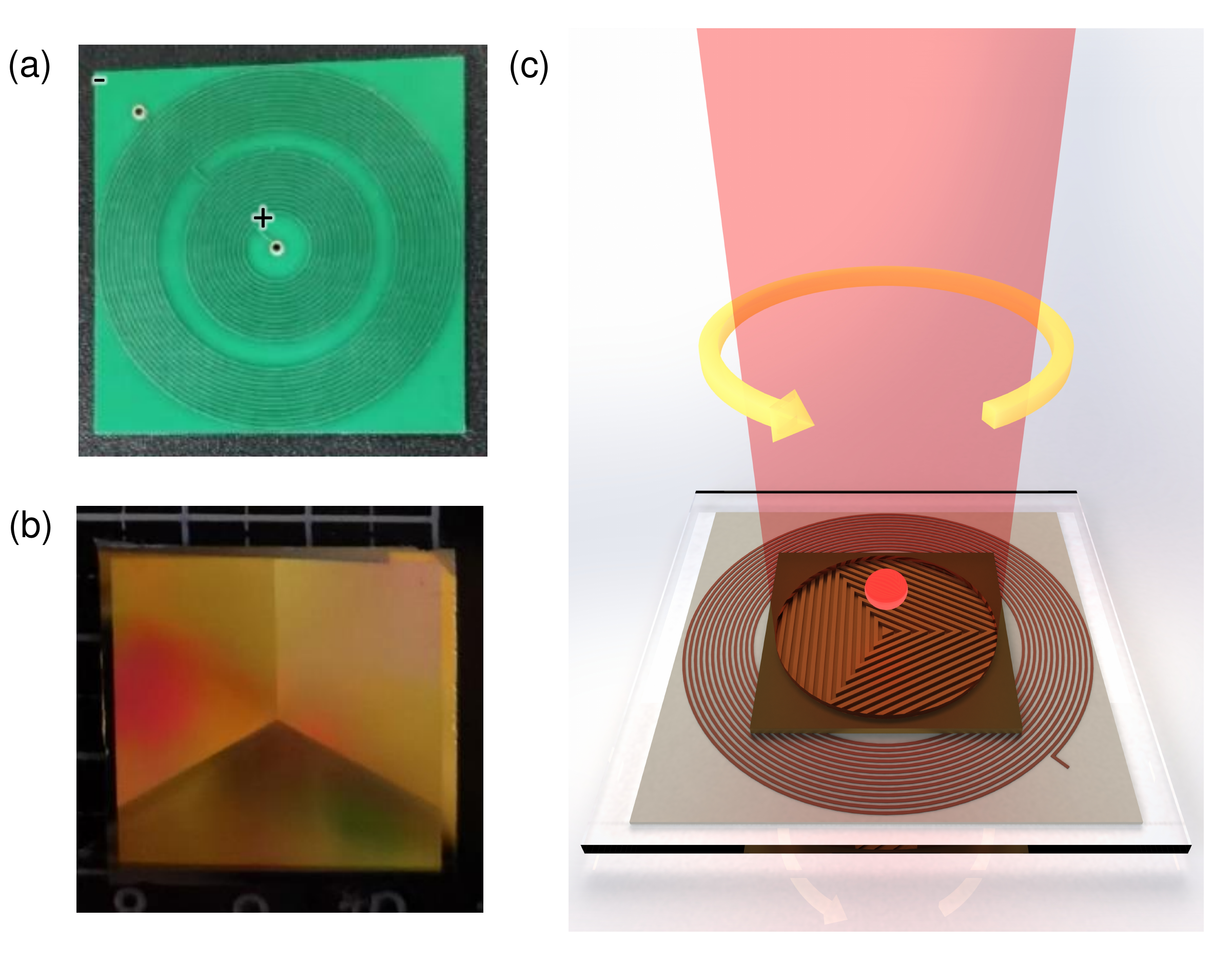}
\par\end{centering}
\caption{Experimental apparatus of the planar-integrated MOT. (a) The coil
chip. (b) The grating chip. (c) Conceptual sketch of the planar-integrated
MOT. The right-handed incident laser beam propagates perpendicular
to the chip and produces three first-order diffracted beams.}

\label{Fig2}
\end{figure}

Figure$\,$\ref{Fig2}(c) shows a conceptual sketch of the co-integrated
MOT based on the combination of a grating chip {[}Fig.$\,$\ref{Fig2}(b){]}
and a coil chip {[}Fig.$\,$\ref{Fig2}(a){]}. The grating chip consists
of three etched gratings on a silicon substrate. Following the pioneering
works of the grating MOT~\citep{Nshii2013,Cotter2016}, the angle
between grating periodic directions is $120{^\circ}$, and the grating
period and duty cycle are $1.42\,\mathrm{\mu m}$ and $0.528$, respectively.
The top side of the grating chip is gold-coated to increase diffraction
efficiency. The measured diffraction angle is $33{^\circ}$ with respect
to $z$-axis and the diffraction efficiency is $40\%$.

In our experiment, the grating chip is mounted underneath and outside
the vacuum cell. The coil chip is mounted below the grating chip and
aligned so that the two working points of the chips coincide. Two
ECDL diode lasers are tuned to the $780\,\text{nm}$ $^{87}\text{Rb}$
D2 lines. The cooling laser frequency is tuned to the $^{5}\text{S}_{1/2}(F=2)$
to $^{5}\text{P}_{3/2}(F'=3)$ cycling transition with a detuning
$\sim12\,\mathrm{MHz}$, and the repump laser is tuned to the $^{5}\text{S}_{1/2}(F=1)$
to $^{5}\text{P}_{3/2}(F=2)$ transition. The beams of the two lasers
are combined and coupled into a polarization maintaining fiber and
delivered to the MOT setup. The incident beam is circularly polarized,
has a diameter of $3.0\,\text{cm}$, an optical intensity of $10\,\mathrm{mW}/\text{cm}^{2}$
for the cooling laser and $1\,\mathrm{mW/}\text{cm}^{2}$ for the
repump. Three first-order diffraction beams from the gratings along
with the incident beam form a MOT in the presence of a quadrupole
magnetic field {[}see Fig.$\,$\ref{Fig2}(a){]}.

\begin{figure}
\begin{centering}
\includegraphics[width=1\linewidth]{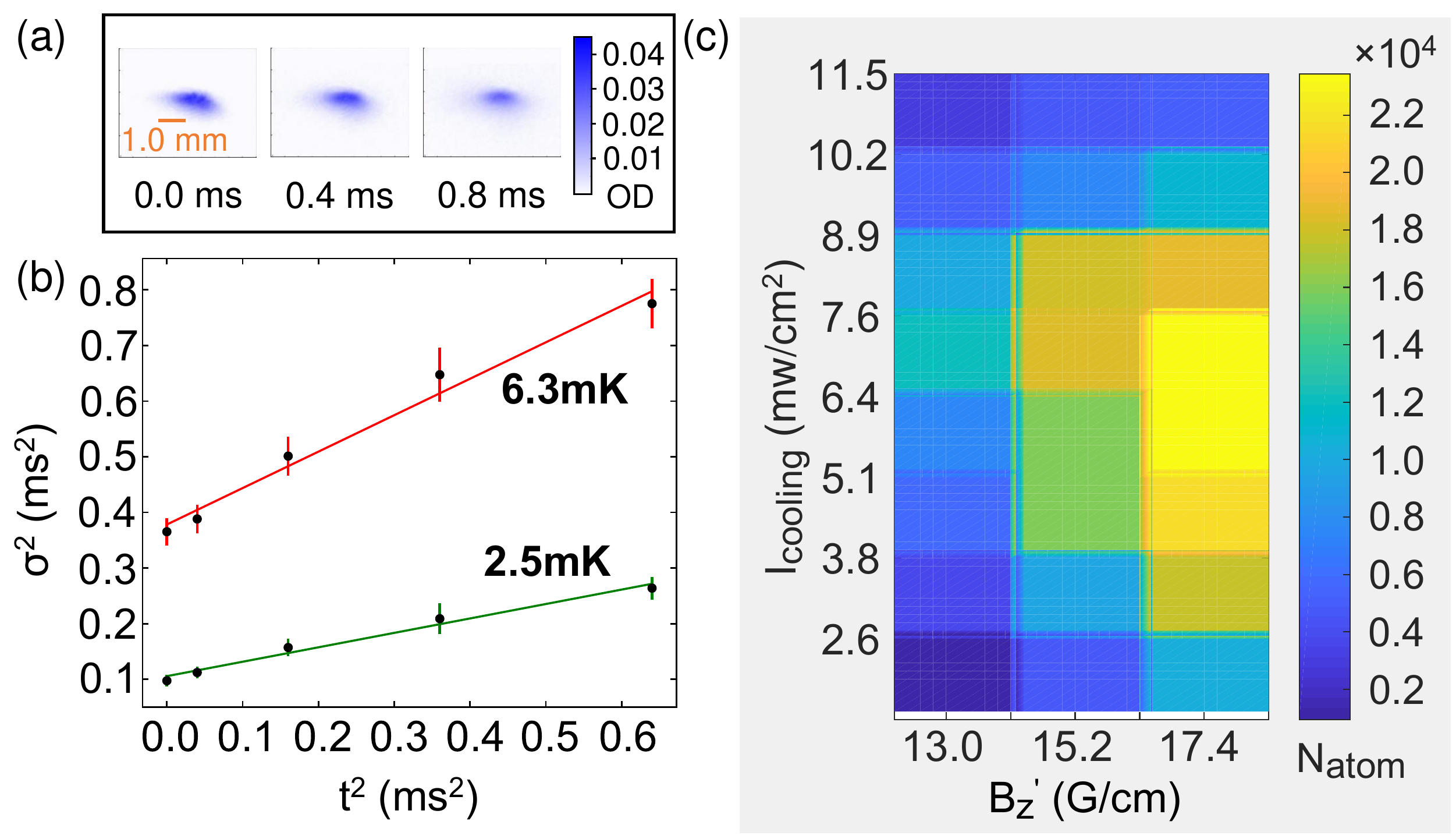}
\par\end{centering}
\caption{Experimental performance of planar-integrated MOT. (a) Several absorption
images of MOT atoms after free expansion over different times $t$.
(b) The fitting result of temperature. Two lines are the fits of $\sigma^{2}$=$\sigma_{0}^{2}$+$k_{B}Tt^{2}/m$
to the data, where $\sigma$ is the 1/e radii of the cloud, $k_{B}$
is the Boltzmann constant, $T$ is the temperature, and $m$ is the
mass of an $^{87}\text{Rb}$ atom. (c) Atom number versus axial magnetic
field gradient and input cooling laser intensity.}

\label{Fig3}
\end{figure}

By absorption imaging and time-of-flight method, the temperature of
the trapped atom cloud is evaluated. Figure$\,$\ref{Fig3}(a) shows
the absorption images of cold atoms following free expansion over
a duration $t$. Figure.$\,$\ref{Fig3}(b) shows the experimental
and fitting results of the atom cloud sizes against $t$. The extracted
temperature is $2.5\,\mathrm{mK}$ for the axial direction and $6.3\,\mathrm{mK}$
for the radial direction. Moreover, we adjust the cooling laser intensity
and the magnetic field gradient to examine the system performance,
and the results are summarized in Fig.$\,$\ref{Fig3}(c). At a trapping
laser intensity of $7\:\mathrm{mW/cm}^{2}$, the planar-integrated
MOT achieves an atom number of $2\times10^{4}$ and a number density
of $6\times10^{6}\:\mathrm{cm}^{-3}$.

\begin{figure}
\begin{centering}
\includegraphics[width=1\linewidth]{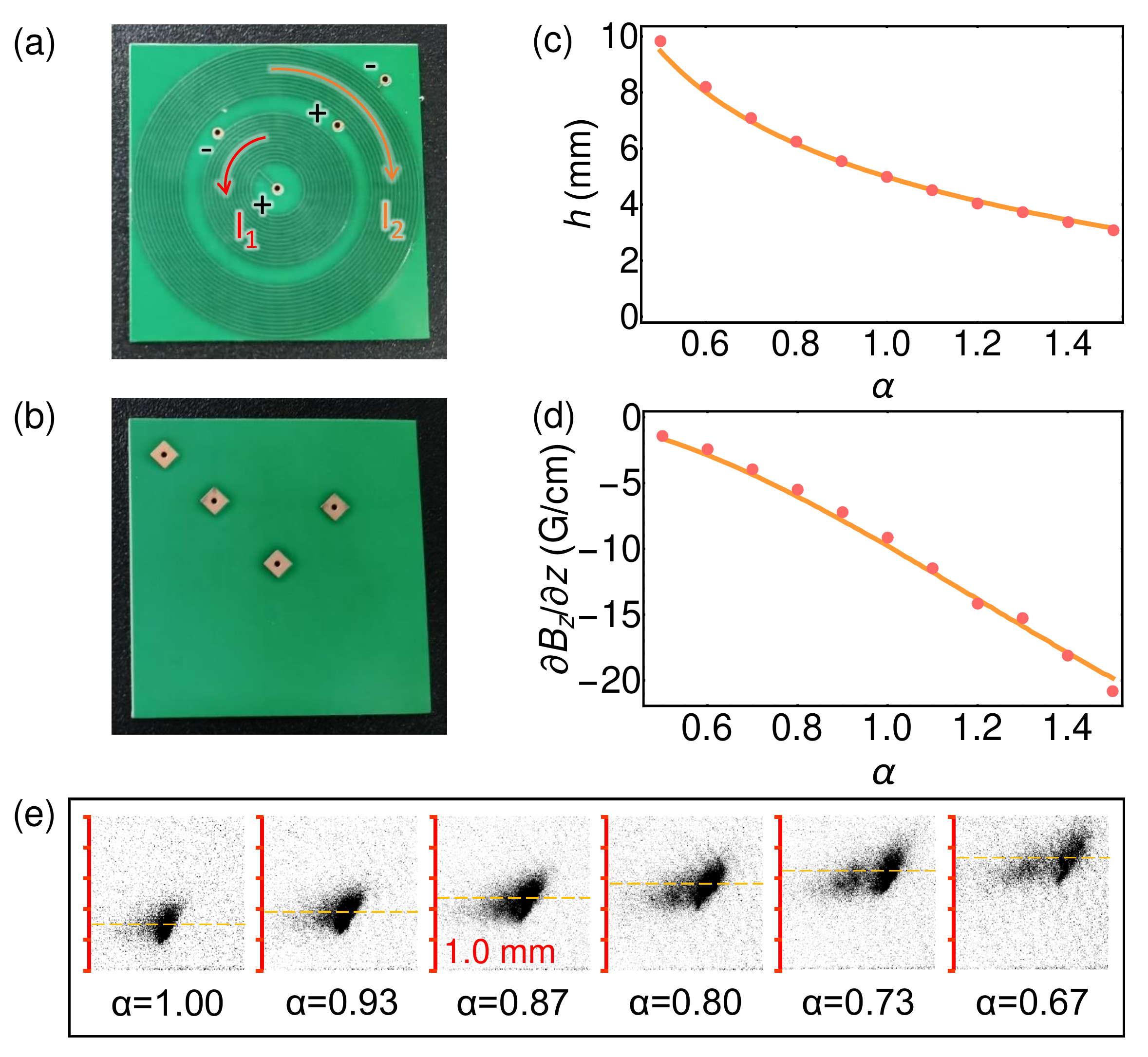}
\par\end{centering}
\caption{Experimental performance of adjusting feature without moving. (a),
(b) Pictures of two sides of another coil chip design with two independent
current supply interfaces. (c) Working point height $z_{\text{0}}$
versus current ratio $\alpha$. (d) Magnetic field gradient at the
working point $B_{\text{z}}^{'}$ versus current ratio $\alpha$.
(e) Several pictures of MOT status under different $\alpha$.}

\label{Fig4}
\end{figure}

According to Eq.$\,$(\ref{eq:1}) and Eq.$\,$(\ref{eq:3}), the
working point moves vertically when adjusting the ratio of the currents
in two planar spirals $\alpha=I_{2}/I_{1}$. The orange lines in Figs.$\,$\ref{Fig4}(c)
and (d) show the simulation results of the height of working point
$h$ and the gradient $\partial B_{z}/\partial z$, with $\alpha$
ranges from $0.5$ to $1.5$ and $I_{1}=0.9\,\mathrm{A}$ is fixed.
We find that the $h$ varies from $9.5\,\mathrm{mm}$ to $3.2\,\mathrm{mm}$,
with the field gradient varying from $1.6\,\mathrm{G/cm}$ to $20\,\mathrm{G/cm}$.
To verify this result, we modified the coil chip with two sets of
printed pads to individually supply the currents for inner and outer
coils. The experimental results (red dots) in Figs.$\,$\ref{Fig4}(c)
and (d) fit well with the theoretical predictions. Then, with such
a modified coil chip, we realized the adjusting of the atom cloud
height from $\sim1.5\,\mathrm{mm}$ to $\sim3.6\,\mathrm{mm}$ by
varying $\alpha$ from $0.67$ to $1.0$, as shown by Fig.$\,$\ref{Fig4}(e).
These results indicate a wide tunability of the working point for
the planar-integrated MOT with future monolithic integration of the
grating and coils.

\section{Conclusion}

In conclusion, we have demonstrated a planar coil configuration to
provide a quadrupole magnetic field for realizing a chip-based MOT.
Combining two chips, for coil and grating, the planar-integrated MOT
is realized for the first time, with significantly reduced volume
and weight. This configuration allows more optical access and reduces
the power requirements on the current supply and heat dissipation.
The planar-integrated MOT also shows excellent compatibility with
photonic chips and convenience in alignments, providing an important
solution towards a fullyintegrated cold-atom system for sensors and
quantum devices.

\smallskip{}

\noindent \textbf{Acknowledgments}\\We would like to thank Tianchu
Li, Yang Yang and Xiaochi Liu for helpful discussions. The work was
supported by the National Key Research and Development Program of
China (Grant Nos. 2016YFA0302200 and 2016YFA0301303), the National
Natural Science Foundation of China (Grant Nos. 11922411, 11874342,
and 41727901), Anhui Initiative in Quantum Information Technologies
(Grant No.~AHY130200), and the Fundamental Research Funds for the
Central Universities (Grant No. WK2470000031). This work was partially
carried out at the USTC Center for Micro and Nanoscale Research and
Fabrication.

\bibliography{reference}

\end{document}